\documentclass[twoside]{hs04}
\usepackage{amsmath,amssymb}
\usepackage{euscript}
\usepackage{cite}

\def\be{\begin{equation}}
\def\ee{\end{equation}}
\def\bc{\begin{center}}
\def\ec{\end{center}}
\newcommand{\bea}{\begin{eqnarray}}
\newcommand{\eea}{\end{eqnarray}}

\def\vh{\varphi}

\def\runauthor{V. Pervushin, V. Zinchuk,
 A. Zorin}
\def\shorttitle{CONFORMAL RELATIVITY: THEORY \& OBSERVATIONS}
\begin{document}

\title{CONFORMAL RELATIVITY: THEORY \& OBSERVATIONS}

\author{V. Pervushin, V. Zinchuk,
 A. Zorin}
\address{Bogoliubov Laboratory of Theoretical Physics,\\
Joint Institute for Nuclear Research, 141980 Dubna, Russia}

\maketitle

\abstracts{Theoretical and observational  arguments
 are listed in favor  of a new principle of
  relativity of  units of measurements as the basis of
 a conformal-invariant
 unification of General Relativity and Standard Model by
  replacement of all masses   with a  scalar (dilaton)   field.
  The relative units  mean conformal observables:
   the coordinate distance, conformal time,
   running masses, and constant temperature. They  reveal to us a
    motion of a universe along its hypersurface in the
   field space of events like a motion of a relativistic
   particle in the Minkowski space, where  the postulate of
   the vacuum as a state with minimal energy leads to arrow of
   the geometric time. In relative units,
  the unified theory describes  the Cold Universe Scenario,
  where the role of the conformal dark energy
  is played by a free  minimal coupling scalar field in agreement
  with   the most recent distance-redshift  data from
type Ia supernovae.
  In this Scenario, the evolution of the Universe begins
  with the  effect of intensive
 creation of   primordial W-Z-bosons  explaining
  the value of CMBR temperature,
 baryon asymmetry, tremendous deficit of the luminosity masses in
 the COMA-type superclusters and large-scale structure of the Universe.}

\section{Introduction}

 Unification of General Relativity (GR) and Standard
Model (SM) is one of  the last bastions of theoretical physics not
won in the 20th century. Our idea of such a unification is  a new
principle of relativity - a relativity of  units of measurements.
This means that equations of motion become  conformal-invariant
ones, they do not depend not only on the data but also on the
units of measurement of these data. In order to obtain the
corresponding conformal unified theory  (i.e., Conformal
Relativity), it is sufficient to replace all masses, including the
Planck mass in GR and Higgs mass in SM, with a scalar massless
dilaton field \cite{plb,kl,pr,ppgc,039}.
 It was shown that this Conformal Relativity
  takes the form of a relativistic
 brane where the dilaton field plays the role of
 the evolution parameter in the field space of events \cite{pr,ppgc,039}.
 Conformal Relativity contains the conformal cosmology
 defined as a version of the standard cosmology where observables
 are identified with the conformal quantities (the coordinate distance,
 conformal time, running masses, and constant temperature).
 Due to this identification the role of the dark energy
 is played by an additional  scalar field (conformal quintessence)
 with the stiff  equation of state in agreement
 with observational data \cite{039,acta,yaf}.

In this paper, we show  how Conformal Relativity can describe
numerous astrophysical data and answer the topical questions of
 modern cosmology.

\section{Conformal-invariant theory}

The first physical theory formulated by Newton was a
representation of the Galilei  group of automorphisms of the
initial data for the Copernicus relativity of the position and
velocity of the Earth. The next step was the Poincar\'e  group
that means the Einstein relativity of the Lorentz frames of
reference \cite{poi,ein} given in the Minkowskian space of events
$[X_0|X_i]$, where a particle in each frame is described by two
times: the time as a variable $X_0$ measured in the rest frame and
the time as a geometric interval $ds$ considered as a measure of
the world line
 $[X_0(s)|X_i(s)]$ of a particle (see Fig.1). An additional datum
 $X_0(s=0)=X_{0I}$
 was treated as the point of creation of a particle moving
 forward, or annihilation of a particle moving backward with
 the postulate of the vacuum as a state with minimal energy \label{pos}defined as
 the canonical momentum $P_0$ of the time-like variable $X_0$.

\vspace{-.51cm}

 \begin{figure}[!hbt]
  \begin{center}
  \includegraphics[width=8cm,height=4cm]{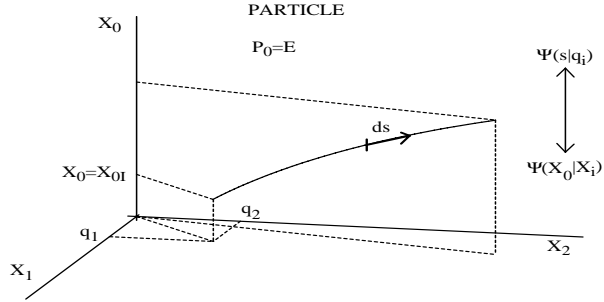}\nopagebreak
  \end{center}
  \caption{The world line $[s]$ of a relativistic particle in the
  space of events [$X_0| X_i$] with the initial data [$X_{0I}| q_I$]
  treated as the point of creation of the particle.}
  \vspace{0.5cm} \label{part}
 \end{figure}

 \vspace{-1cm}

 The  Poincar\'e group of automorphisms of
 the initial data follows from symmetries of the Faraday -- Maxwell
 electrodynamics that has one more symmetry -- conformal \cite{bat,w49}.
 Conformal-invariant unified theories equivalent to GR
  and SM have been proposed by a lot of
 authors (see, for example \cite{plb,kl}, and references therein), who
 replaced all masses with the scalar dilation field $w$. One of such
 theories has
 the action  \cite{ppgc}
 \be\label{t1}
 S=S_{\rm GR}+S_{\rm SM}+
 \int d^3x \sqrt{-g}w^2\partial_{\mu}Q\partial^{\mu}Q,
 \ee
where $
 S_{\rm GR}=
 -\int d^4x\left[\sqrt{- g}w^2 {}^{(4)}R(g)/6-
 w\partial_\mu\left(\sqrt{- g}\partial^\mu w\right)\right]
 $
is the Penrose -- Chernikov -- Tagirov action \cite{pct}, $S_{\rm
SM}$ is the Standard Model in that the Higgs mass is replaced with
the dilaton $y_hw$ ($y_h\sim M_h/M_{Planck}\simeq 10^{-17}$), and
the last term is included in the theory as one of the possible
models of the dark energy.

Two dilatons $X_0=w\cosh Q$ and $X_1=w\sinh Q$  with signature
($+, -$)
 and four Higgs fields
 $|\Phi_{\rm Higgs}|^2=X_2^2+X^2_3+X_4^2+X_5^2$ give a
 scalar sector of the action (\ref{t1}) in the form of a relativistic
 Brane (with the metric $G^{AB}={\rm diag}(+1,-1,-1,-1,-1,-1)$):
\be \label{br} S_{\mbox{\scriptsize BRANE}}= -\int
d^4x\left[\sum\limits_{A,B=0}^{5}G^{AB}\left[\sqrt{-g}X_AX_B\frac{R}{6}-
X_A\partial_\mu\left( \sqrt{-g} g^{\mu\nu}\partial_\nu
X_B\right)\right] \right]
 \ee
 that is analogous to the action of a relativistic particle
\be \label{sr}
 S_{\mbox{\scriptsize PARTICLE}}=-\frac{m}{2}\int\limits_{x^0_1}^{x^0_2}dx^0
\left[\frac{1}{\bar e} \left(\frac{d X_0}{dx^0} \frac{d
X_0}{dx^0}-\frac{d X_i}{dx^0} \frac{d X_i}{dx^0}\right)+{\bar
e}\right]. \ee

The Conformal
 Relativity (\ref{t1}) considered above can be easily
 converted into the conventional  GR and SM
 by a scale transformation, so that the new dilaton
 field $\widetilde{w}$
 coincides with its occasional present-day ``datum'' \cite{kl}
 \be \label{pt}
  \widetilde{w}(x)=\vh_0=M_{Planck}\sqrt{3/8\pi}~~~~~~~~~~~~
  (c=1,~\hbar=1).
  \ee
 The defect of this choice of
 variables is the conversion of  an occasional datum into
 the  fundamental parameter of equations of
 motion\footnote{This  fixation of the dilaton date $\vh_0$
   looks like the  Ptolemaeus absolutization
   of the Earth ``initial datum'' in the Newton
 mechanics.}.

\vspace{-.51cm}

 \begin{figure}[!hbt]
  \begin{center}
  \includegraphics[width=8cm,height=5cm]{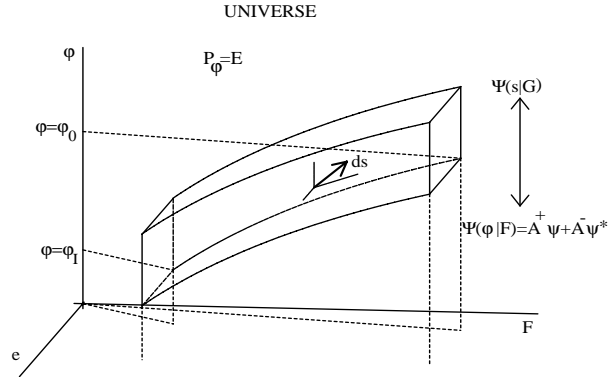}\nopagebreak
  \end{center}
  \caption{The world hypersurface [$s$] of the Universe in the field space of
  events [$\varphi|F$] with the vacuum initial data
  [$\varphi_I|G$] treated as the point of creation of the Universe.}
  \vspace{0.5cm} \label{uni}
 \end{figure}

\vspace{-1cm}

 Another choice of variables in  Conformal
Relativity (\ref{t1}) is given  by   analogy of
 with the theory of a relativistic particle (\ref{sr}) mentioned above.
 In a concrete frame of reference
 both the theories are invariant
 with respect to the reparametrizations of
 the coordinate ``time''
 $x^0\to \widetilde{x}^0=\widetilde{x}^0(x^0)$. This invariance,
 in theories of all relativistic systems (particle \cite{poi,ein},
 string \cite{rf}  and General Relativity \cite{pp}), means
 that this  coordinate ``time'' is not observable.
 As we have seen above, there are two observable ``times'':
 i) a ``time'' as a time-like dynamic variable in the field space of
 events, and
 ii) a ``time'' as an invariant geometric interval.
 Conformal symmetry allows us to choose
  such a time-like dynamic variable as the dilaton
 \be\label{rel}w(x^0,x^i)=\vh(x^0)
 \ee
  because it has the negative
 sign of its action.  This choice
 (in contrast to naive fixation (\ref{pt}))
 keeps   the relative
 units, and it   reveals to us a
 motion\footnote{The dilaton canonical momentum $P_\vh$
  is considered as an energy $E$ of events. Stability of a
  relativistic theory requires
  the postulate of the vacuum as a state with minimal energy $E$.
The vacuum postulate restricts the motion of the Universe in the
field space of events and it
 means that for positive energy of events
  the Universe moves forward $\vh>\vh_I$,
 and the anti-Universe moves backward $\vh<\vh_I$,
 where $\vh_I$ is the initial data treated in quantum theory
 as a point of creation, or annihilation, respectively.
One can see that  a universe with the positive energy of events
does not contain the cosmological singularity $\vh =0$ that
 belongs to an anti-universe.} of a
  relativistic universe along the world hypersurface in the field space of
  events $[\vh|F]$ from a point $\vh_I$ to the present-day
  point $\vh_0$ defined by Eq. (\ref{pt}))
  (see Fig.2).

\section{Conformal observables}

 The relative units  (\ref{rel}) mean that the dilaton as a
 cosmological scale factor $a=\vh/\vh_0=1/(1+z)$ scales all masses.
 In the Conformal Relativity
 the conformal version of Friedmann cosmology   arises
 without any assumption about homogeneity as averaging of an
 exact equation over the spatial volume (see Appendix A).
  In the conformal cosmology
 observational quantities are identified
 with the conformal ones: conformal time $d\eta$ (instead of the Friedmann
 one $dt=a(\eta)d\eta)$,
 coordinate distance $r$ (instead of
  $R= r/(1+z)$),
 running masses $m=m_0/(1+z)$ (instead of the constant one
 $m_0$), and the conformal temperature $T_c(z)=T(z)/(1+z)$
 (instead of the standard one $T(z)$). In this case
 the
  red shift of the spectral lines of atoms on cosmic objects
 $$
\frac{E_{\rm emission}}{E_0}=\frac{m_{\rm atom}(\eta_0-r)}{m_{\rm
atom}(\eta_0)}\equiv\frac{\vh(\eta_0-r)}{\vh_0}=a(\eta_0-r)
=\frac{1}{1+z}
$$
is explained by the running masses. The conformal observable
distance  $r$ loses the factor $a$, in comparison with the
nonconformal one $R=ar$, therefore, the  recent distance-redshift
data from type Ia supernovae \cite{SN},
 in the conformal
 cosmological model \cite{039}, are compatible with the stiff state
 $p=\rho$ of the conformal
 quintessence (\ref{t1}) and the square root dependence of
 the scale factor on the conformal (i.e., observable) time $a(\eta)\sim \sqrt{\eta}$;
 this dependence  can explain chemical evolution \cite{three}.

 As it has been shown in
\cite{ppgc,yaf}, when the Universe horizon coincides with the
Compton length of the vector bosons $H_I=M_{I}$,  there is an
intensive cosmological creation of the primordial vector bosons
from the vacuum (see Fig.3).  This creation leads to a conformal
temperature in the form of the integral of
 motion of the Universe in the  stiff state
 $T_I=(M_I^2H_I)^{1/3}$  as a consequence of
 collision and scattering of these bosons.

\begin{figure}
  \centering
  \includegraphics{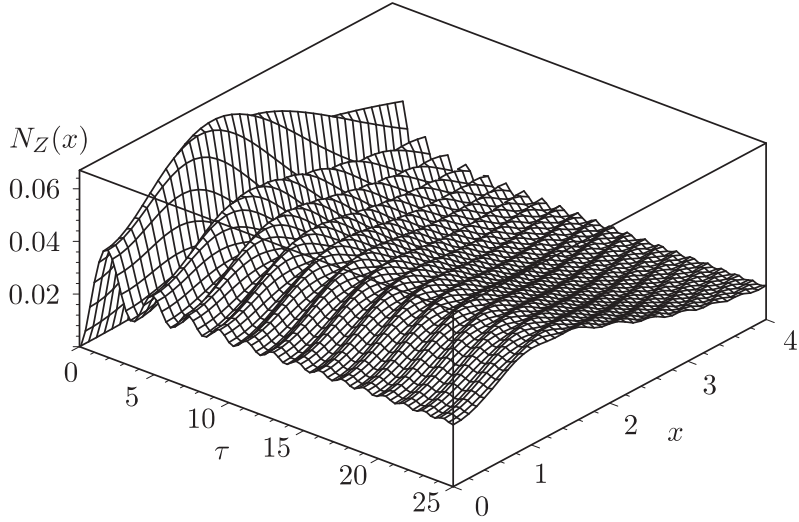}
  \caption{Longitudinal ($N_Z(x)$) components
of the boson distribution versus the dimensionless time $\tau=
2\eta H_I$ and the dimensionless momentum $x = q/M_I$ at the
initial data $M_I = H_I$ ($\gamma_v = 1$).}
\end{figure}

  These theoretical
results are  in satisfactory agreement with the value of
temperature of Cosmic Microwave Background radiation as a final
product of decays of the primordial bosons remembering the
integral of motion \be\label{t}T_{\rm
CMB}\sim(M_I^2H_I)^{1/3}=(M_{W0}^2H_0)^{1/3}\sim 3 K.\ee The
 value of the baryon--antibaryon asymmetry
of the Universe is followed from the axial anomaly and is frozen
by the superweak-interaction with the coupling constant
 \be\label{a}X_{CP}=n_b/n_\gamma\sim 10^{-9}.
 \ee The boson
life-times    \cite{yaf} $\tau_W=2H_I\eta_W\simeq
\left(\frac{2}{\alpha_g}\right)^{2/3}\simeq 16,~ \tau_Z\sim
2^{2/3}\tau_W\sim 25
 $ determine the present-day visible
baryon density
\be\label{b}\Omega_b\sim\alpha_g=\alpha_{QED}/\sin^2\theta_W\sim0.03.\ee

 This baryon density as a final product of the  decay of bosons
 with momentum $q$ and
 energy $\omega(\eta)=(M^2(\eta)+q^2)^{1/2}$ oscillates as
 $\cos\left[2\int_0^{\eta}d\bar \eta \omega(\bar \eta) \right]$ \cite{yaf}.
 One can see  \cite{vin} that the number  of  density oscillations of the
 primordial bosons during their life-time for the momentum $q\sim
 M_I$  is of order of $20$, which is very close to the number of
 oscillations of the visible baryon matter density recently
 discovered in researches of large scale periodicity in redshift
 distribution \cite{a1,a2}
 \be\label{ls}
 [H_0\times128~{\rm MPc}]^{-1}\sim 20\div25\sim (\alpha_g)^{-1}.
 \ee
  The results (\ref{t}), (\ref{a}), (\ref{b}), (\ref{ls})
 testify to that all  visible matter can be a product of
 decays of primordial bosons with the oscillations forming
 a large-scale structure of the baryonic matter. The number $N_s$
 of  superclusters can be estimated from Eq. (\ref{ls}):
 $N_s\sim (\alpha_g)^{-3}$.
  Each of them has gravitation radius  of  order of $r_g\sim
 (\alpha_g)^{4}H_0^{-1}$.

 Cosmological perturbation theory \cite{480} in Conformal Relativity
 (in contrast to the standard one \cite{lif}) does not contain
 the scalar metric
 component
 as a dynamic variable and contains the Yukawa-like
  form of interactions with the
shift  of the coordinate
  origin by a central gravitation field (see Eqs.  (\ref{h}) -- (\ref{in})).
  Due to this shift the field of each supercluster can lead to the spatial
 anisotropy of fluctuations of the photon energy (\ref{f})
 and temperature of CMBR: $${\triangle
T}/{T}\sim |r'_g|=r_g\times H_0\sim\alpha^4_g\sim
10^{-5}\div10^{-6}.$$

In Conformal Relativity, galaxies and their  clusters are formed
by the Newton Hamiltonian with running masses
$E(\eta)={p^2}/{2m(\eta)}-{r_g(\eta)m(\eta)}/2r$, where the Newton
coupling ${r_g(\eta)m(\eta)}/{2}=r_g(\eta_0)m(\eta_0)/2$ is a
motion constant. One can see that the running masses lead to the
effect of the capture of an object by a gravitational central
field at the time when $E(\eta_{\rm capture})=0$. After the
capture the conformal size of the circle trajectories decreases as
${r(\eta)=R_{\rm circle}/a(\eta),~R_{\rm circle}=\rm const}$.

 In the stiff state,
the running masses change the orbital curvatures \cite{dm}
 $$v_{\rm orbital}(R_{\rm circle})=\sqrt{\frac{r_g}{2R_{\rm circle}}+2(R_{\rm
circle}H)^2}$$ which  can explain the tremendous deficit of the
luminous matter $M/M_{L}\sim 10^2$, where $M_L$ stands for the
mass of luminous matter, in superclusters with a mass $M\geq
10^{15}M_\odot$, $R\gtrsim 5 {\rm Mpc}$\cite{td}, where the Newton
velocity becomes less than the cosmic one.

\section*{Conclusion}

 We listed arguments in favor of
 that the Conformal Relativity can explain
problems of energy, arrow of time,  cosmological singularity,
homogeneity, origin of all visible matter from physical vacuum,
CMBR conformal temperature, baryon asymmetry, large-scale
structure of the universe expressed in terms of  fundamental
parameters of the SM. The listed arguments stimulate  farther
consideration  of other topical problems of modern cosmology,
including gravitational lensing, and fluctuation of the CMBR
temperature, in terms of relative units.

\section*{Acknowledgements} We thank B. Barbashov, D. Blaschke,
S. Dubni\'cka, A. Gusev, E. Kuraev, L. Lipatov, S. Vinitsky, and
A. Zakharov for discussion. One of the authors (V.P.) thanks Prof.
A. Dubnichkova for the hospitality in Comenius University of
Bratislava.
 This work was
supported in part by the Votruba - Blokhintsev Grant.

\section*{Appendix A: Hamiltonian approach to Conformal Relativity}

\renewcommand{\theequation}{A.\arabic{equation}}

\setcounter{equation}{0}

  The Hamiltonian approach to Conformal Relativity (\ref{t1})
  in the relative units  (\ref{rel}) is formulated
  in a frame of reference  given by a geometric interval
 \be
 \label{ds}
 g_{\mu\nu}dx^\mu dx^\nu\equiv\omega_{(0)}\omega_{(0)}-
 \omega_{(1)}\omega_{(1)}-\omega_{(2)}\omega_{(2)}-\omega_{(3)}\omega_{(3)}
 \ee
 where  $\omega_{(\alpha)}$ are  linear differential forms  \cite{fock29} in
 terms of the Dirac variables \cite{dir}
\be \label{adm}
 \omega_{(0)}=\psi^6N_{\rm d}dx^0,~~~~~~~~~~~
 \omega_{(a)}=\psi^2 {\bf e}_{(a)i}(dx^i+N^i dx^0);
 \ee
 here triads ${\bf e_{(a)i}}$ form the spatial metrics with $\det |{\bf
 e}|=1$.
 These forms are invariant with respect to
 the kinemetric general coordinate transformations \cite{vlad} $\widetilde{x}^0=\widetilde{x}^0(x^0),
 \widetilde{x}^i=\widetilde{x}^i(x^0,x^l)$. This invariance means that the frame
 of reference (\ref{adm}) should be redefined by two {``times''}: the
 {``time''} as a variable $\vh$ in the field space of events
 $[\vh|F]$ (see Fig. 2) and the {``time''} as a geometric
 interval \cite{rf, pp}
 \be\label{ght} d\zeta=N_0(x^0)dx^0; ~~~~~~
 ~\zeta(x^0)=\int\limits_{}^{x^0}d\overline{x}^0N_0(\overline{x}^0),
 \ee
 where
\be \label{n0}
 {N_0(x^0)}^{-1}={V^{-1}_{0}}\int_{V_0}d^3x{ N^{-1}_d(x^0, x^i)}
 \equiv\left\langle{ N^{-1}_d}\right\rangle%
 \ee
 is the averaging of the  inverse  lapse
 function $N_d$ over  spatial volume $V_0=\int d^3 x$.

The  Hamiltonian action  in the field space $[\vh|F]$ takes the
form \cite{480}
 \be\label{12ha1}
 S=\int dx^0\left[-P_{\vh}\partial_0\vh+
 N_0\frac{P^2_\vh}{4V_0}+\int d^3x
 \left(\sum\limits_{{F}
 } P_{F}\partial_0F
 +{\cal C}-N_d \psi^{12}{T^0_0}_t\right)\right],
 \ee
 where $P_{ F}$ is the set of the field momenta
 $p_{\psi}, p^i_{{(a)}},p_f,p_{Q}$;
 the sum of constraints
 \be\label{2ha3}
 {\cal C}=N^i {T}^0_{i} +C_0p_\psi+ C_{(a)}\partial_k{\bf e}^k_{(a)}
 \ee
contains  the weak Dirac constraints\footnote{We impose the strong
constraint $\int_{V_0}d^3 x {p}_\psi\equiv 0$ to remove the double
counting of the spatial metric determinant and keep the number of
variables of GR (in contrast with the standard cosmological
perturbation theory \cite{lif}).}
 of transversality and the minimal space-like surface \cite{dir}
 \be\label{gauge}
 \partial_i {\bf e}^{i}_{(a)}\simeq 0;~~~~p_{\psi}=\frac{8\vh^2}{N_d}\left[
 (\partial_0-N^l\partial_l)\ln{\psi}-\frac16\partial_lN^l\right]\simeq 0,
 \ee
 respectively, with the Lagrangian multipliers $C_0,~C_{(a)}$;
 \bea\label{2ha4}
 {T^0_0}_t&=&\frac{1}{\psi^{12}}\left[\frac{6p_{(ab)}p_{(ab)}}{\vh^2}
 -\frac{16}{\vh^2}p^2_{\psi}\right]
 +\frac{\varphi^2}{6\psi^{5}}~
 \left[{}^{(3)}R({\bf e})\psi+
 8\triangle\psi\right]+T^0_{0({\rm SM})},\\\label{t0k}
 {T^0_k}_t&=& -p_{\psi}\partial_k \psi+\frac{1}{6}\partial_k
 (p_{\psi}\psi) +p^i_{(b)}\partial_k{\bf e}_{i(b)}+ T^0_{k({\rm SM})}~
 \eea
 are the total components of the energy  - momentum tensor
 in terms of momenta
 $p_{(ab)}=\left[p^i_{(a)}{\bf e}_{(b)i}+p^i_{(b)}{\bf
 e}_{(a)i}\right]/2$.

 The gauge-invariant lapse function $N_d/N_0={\cal N}$ and
   the spatial metric determinant $\psi$ can be determined by
   their equations for both the zero Fourier harmonic
   $\langle F\rangle$ and the nonzero ones $\overline{F}=F-\langle
   F\rangle$:
 \begin{align}\label{f0}
 \left\langle N_d\frac{\delta S[\vh]}{\delta  N_d }\right\rangle=0&
 |\!\!\models\!\!\!\!\!\!
 \Longrightarrow\vh'^2=\rho_t,&
 \overline{N_d\frac{\delta S[\vh]}{\delta  N_d}}=0~
 |\!\!\models\!\!\!\!\!\!&\Longrightarrow
 \frac{\rho_t}{\cal N}={\cal N}{\cal H}_t, \\
 \left\langle
 \psi\frac{\delta S[\vh]}{2\delta \psi}\right\rangle
 =0&
 |\!\!\models\!\!\!\!\!\!
 \Longrightarrow(\vh^2)''=3(\rho_t-p_t),&
 \overline{\psi\frac{\delta S[\vh]}{2\delta \psi}}
 =0~
 |\!\!\models\!\!\!\!\!\!&
 \Longrightarrow~ \overline{\hat{\bf A}{\cal N}}=0
 \label{4f2},
 \end{align}
 where  $\varphi'\equiv{d\varphi}/{d\zeta}$,
 $\rho_t\equiv\langle{\cal N}{\cal H}_t\rangle=
 \langle{\cal N}\psi^{12}{T^0_0}_t\rangle$ and
 $p_t=\langle{\cal N}\psi^{12}{T^k_k}_t\rangle/3$
 are the energy density and pressure of all fields,
 respectively; ${\cal H}_t=\psi^{12}{T^0_0}_t$, and
 $\hat {\bf A}$ is a differential operator defined by identity:
 $$
 \hat {\bf A} {\cal N}\equiv\frac{2\varphi^2}{3}
 \left[({}^{(3)}R({\bf e})\psi^8+
 8\psi^{7}\triangle\psi) {\cal N}+
  \partial_j[\psi^2\partial^j (\psi^6
 {\cal N})]\right]+\psi^{12}[3T^0_{0({\rm SM})}-T^k_{k({\rm SM})}]{\cal N}.
 $$
  Eqs. (\ref{f0}), (\ref{4f2}) show us  that their zero harmonics
  coincide with the conformal version of the Friedmann
  equations with the scale factor $a=\vh/\vh_0$.

 The energy constraints (\ref{f0}) have solutions
 $ P_{\vh(\pm)}=\pm 2V_0\vh'=\pm 2V_0{\langle \sqrt{\cal H}_t\rangle}$,
 ${\cal N}={{\langle \sqrt{\cal H}_t\rangle}}/{\sqrt{{\cal
 H}_t}}$.
 If we substitute these solutions into the action (\ref{12ha1}),
  we obtain, in the case of the positive energy of events,
    the reduced Hamiltonian action
 \be\label{2ha2} S_{+}[\varphi_I|\varphi_0]|_{\rm
energy
 ~constraint}
 =
 \int\limits_{\vh_I}^{\vh_0} d\vh\left\{\int d^3x
 \left[\sum\limits_{  F}P_{  F}\partial_\vh F
 +\bar{\cal C}-2\sqrt{{\cal H}_t}\right]\right\},
 \ee where
$\bar{\cal C}={\cal
 C}/\partial_0\vh$ and
 $\vh_I$ is an initial datum.
  The action (\ref{2ha2}) gives
  the evolution of fields directly in terms of the redshift parameter
  connected with the scale factor $\vh$ by the relation $\vh=\vh_0/(1+z)$.
  Using the reduced action (\ref{2ha2}) one can get
   the probability to find the Universe at the point
 $(\vh_I,F_I)$, if the Universe was created at the point $(\vh_0,
 F_0)$  determined by the causal Green function:
  \be\label{gg}
 G(\vh_IF_I|\vh_0 F_0)=G_{+}(\vh_IF_I| \vh_0
 F_0)\theta(\vh_0-\vh_I)+
 G_{+}(\vh_0F_0|\vh_IF_I)\theta(\vh_I-\vh_0), \ee
where $G_{+}$ can be  given by the Faddeev -- Popov (FP)
functional integral
 \be G_{+}(\varphi_IF_I|\varphi_0
 F_0)=\int \prod_{F={\bf e},\psi,f}
 \left(\frac{dP_{F}dF}{2\pi}\right)\delta(T_0^i)\delta(p_\psi)
 \delta(\partial_i{\bf e}^i_{(a)})
 D e^{iS_{+}[\varphi_I|\varphi_0]}, \ee
 here
$D$ is the FP determinant of the matrix
$D_{(b)(a)}F_{(b)}=\partial_{(b)}\partial_{(a)}F_{(b)}$.
 The vacuum postulate (defined like for relativistic particle
 one on page \pageref{pos}) restricts a motion of a
  universe forward $\vh_0 \geq \vh_I$ for a positive energy
  $E=P_\vh=2V_0\vh'\geq 0$, and backward $\vh_0 \leq \vh_I$
  for a negative energy
  $E\leq 0$ treated as annihilation of a universe at the point $\vh_I$,
  so that the geometric time  (\ref{ght}) as a solution of (\ref{f0})
 is always positive
 \be
 \label{111}\zeta(\varphi_0|\varphi_I)=\theta(\vh_0-\vh_I)\int_{\vh_I}^{\vh_0}
 \frac{d\vh}{\sqrt{\rho_t(\varphi)}}+
 \theta(\vh_I-\vh_0)\int_{\vh_0}^{\vh_I}
 \frac{d\vh}{\sqrt{\rho_t(\varphi)}}\geq 0.
 \ee
%%%%%%%%%%%%%%%%%%%%%%%%%%%%%%%%%%%%%%%%%

Conformal Relativity gives a new cosmological perturbation theory
\be\label{ncpt} \omega_{(0)}=(1- \overline{\Phi}_N)d\eta,~~~~
\omega_{(a)}=(1+\overline{\Phi}_h) (dx_{(a)}+h^{(TT)}_{(a)i}dx^i+
\partial_{(a)}\sigma
d\eta+N^{(T)}_{(a)}d\eta),\ee where the metric components are
defined in the class of functions
 with the nonzero Fourier harmonics
$ \widetilde{\Phi}(k)=\int d^3x \overline{\Phi}(x)e^{ikx}$.
 % satisfying the strong constraint
% $\int d^3x \overline{\Phi}(x)\equiv 0$.
 In the approximation
 $\rho_s=\langle T^0_{0\rm (SM)}\rangle\gg
 \overline{T}_{0}^0=(T_{0\rm (SM)}^0-\langle T^0_{0\rm (SM)}\rangle);
 3p_s=\langle T^k_{k\rm (SM)}\rangle \gg\overline{T}^k_{k}
 =(T_{k\rm (SM)}^k-\langle T^k_{k\rm (SM)}\rangle)
 $, the equations of
 the theory
 %where $T_0^0\equiv\langle T^0_0\rangle +(T_0^0-\langle T^0_0\rangle)
 % =\rho_s+\overline{T}_{0}^0$,
 for the scalar, vector,  and tensor components   take form
  \begin{align}\label{107}
\widetilde{T}^0_{0}&=\frac{2\varphi^2k^2}{3}\widetilde{\Phi}_h+
2\rho_s\widetilde{\Phi}_N, \\\label{108}
\widetilde{T}^0_{0}+\widetilde{T}^k_{k}&=
+9(\rho_s-p_s)\widetilde{\Phi}_h+
 \left(\frac{2\varphi^2k^2}{3}+
5\rho_s+3p_s\right)\widetilde{\Phi}_N,\\
  T^{0\:(T)}_k&=-\frac{\varphi^2}{12}N^{T}_{k};~~~
\overline{T}^{TT}_{ik}=\frac{\varphi^2}{12}\left[-\triangle
 h^{(TT)}_{ik}+
 \frac{(\varphi^2{h^{TT}_{ik}}'){\vphantom{h^{TT}_{ik}}}'}{\varphi^2}\right],
 \end{align}
 where
 $ \partial_i T^{(TT)}_{ik}=T^{(TT)}_{ii}=0,
~~ \partial_kT_k^{0\:(T)}=0$. The Dirac minimal surface constraint
(\ref{gauge}) defines the longitudinal shift vector (\ref{ncpt})
$\triangle\sigma=\frac{3}{4}\overline{\Phi}_h'$.
 In the
 Newton case $p_s,
 \rho_s\ll{\varphi^2k^2}$,
 we obtain standard classical solutions with the Newton constant $G$:
 \be\label{dgg}
 \widetilde{\Phi}_h=\frac{4\pi G}{k^2}\widetilde{T}^{0}_{0};~~~~
 \widetilde{\Phi}_N
 =\frac{4\pi G}{k^2}\left[\widetilde{T}^{0}_{0}
 +\widetilde{T}^{k}_{k}\right];~~ G= \frac{3}{8\pi\varphi^2}.
 \ee
  In the case of point mass distribution with the zero pressure
  ${T}_{k}^k=0$ and  the  density
  \be\label{t00}
 \rho_s=\sum_J \frac{M_J}{V_0}=a\Omega_bH_0^2\vh_0^2;~~~~~~~~~~
 ~~\overline{T}_{0}^0=\sum_J M_J\left[\delta^3(x-y_J)-\frac{1}{V_0}\right],
 \ee
 solutions of Eqs. (\ref{107}), (\ref{108}) take
 the form:
 \be\label{h}
  \overline{\Phi}_h(x)=\sum_{J}
  \frac{r_{gJ}}{2r_{J}}\left[1.15e^{\{-\sqrt{11.25}\mu (z)r_{J}\}}
  -0.15\cos[\sqrt{3.75}\mu(z)r_{J}]\right],\ee
 $$%\be\label{n}
 \overline{\Phi}_N(x)=\sum_{J}
 \frac{r_{gJ}}{2r_{J}}\left[0.95e^{\{-\sqrt{11.25}\mu (z)r_{J}\}}+
  0.05\cos[\sqrt{3.75}\mu(z)r_{J}]\right],~~~~~~~
 $$%\ee
 where $r_{gJ}=2GM_J$, $r_{J}=|x-y_J|$,
  and  $\mu(z)=\sqrt{\Omega_b}H_0(1+z)^{1/2}$.
  The minimal surface (\ref{gauge})
 gives the shift of the coordinate
  origin in the process of evolution:
 \be\label{ni}
 N^i=\sum_{J}\frac{3r_{gI}'}{8}
 \frac{(x-y_J)^i}{|x-y_J|}\left[1.15e^{\{-\sqrt{11.25}\mu (z)r_{J}\}}
  -0.15\cos[\sqrt{3.75}\mu(z)r_{J}]\right].
  \ee
  The obtained conformal interval \be\label{in}
 ds^2_c=(1-2\overline{\Phi}_N)d\eta^2-(1+2\overline{\Phi}_h)
 (dx^i+N^id\eta)^2 \ee
  determines an equation for the photon momenta
  \be
  p_{\mu}p_{\nu}g^{\mu\nu}\simeq (p_0+N^ip^i)^2(1+2\overline{\Phi}_N)
  -p^2_j(1-2\overline{\Phi}_h)=0,\ee
 from which we obtain a photon energy
  \be
 p_0\simeq-N^ip^i+[1-({\overline{\Phi}_N+\overline{\Phi}_h})]|p|;~~~~~~~ |p|=\sqrt{p^2_i}. \ee
 This formula shows us
 the relative magnitude of  spatial fluctuations of a photon energy
 in terms of  the metric components (the potential $\overline{\Phi}$ and
 shift vector $N^i$) \be\label{f}
 \frac{p_0-|p|}{|p|}=-[N^in^i+({\overline{\Phi}_N+\overline{\Phi}_h})];~~
 ~~~~~~~n^i=\frac{p_i}{|p|}.
  \ee
 %The influence non-synchronic frame of reference (i.e. presence of
% $N^i$) on propagation of photons in the space with distributed point mass
% object in it (\ref{t00}) leads to
 The appearance of  spatial anisotropic
 fluctuations of the photon energy in the flow of photons is
 the consequence of the minimal surface  (\ref{gauge}) that leads
 to the hermitian Hamiltonian in the field space of events $[\vh|F]$.

The cosmological perturbation theory in Conformal Relativity does
not require its convergence to be proved because the perturbations
are in a different class of functions (with nonzero Fourier
harmonics) than the cosmological dynamics described by the
equations in the zero harmonic sector.

\end{document}